\documentclass[11pt,a4paper]{article}
\usepackage{jcappub}
\hypersetup{pdfstartview=}
\usepackage{hhline}
\usepackage{wasysym}

\usepackage{calligra}

\DeclareMathAlphabet{\mathcalligra}{T1}{calligra}{m}{n}
\DeclareFontShape{T1}{calligra}{m}{n}{<->s*[2.2]callig15}{}
\newcommand{\scriptr}{\mathcalligra{r}\,}

\title{Astrophysical constraints on compact objects in 4D Einstein-Gauss-Bonnet gravity
}

\author[a]{C.~Charmousis,}
\author[b]{A.~Leh\'ebel,}
\author[c]{E.~Smyrniotis,}
\author[c]{N.~Stergioulas}

\affiliation[a]{Université Paris-Saclay, CNRS/IN2P3, IJCLab, 91405 Orsay, France.}
\affiliation[b]{Centro de Astrof\'{\i}sica e Gravita\c c\~ao  - CENTRA,
Departamento de F\'{\i}sica, Instituto Superior T\'ecnico - IST,
Universidade de Lisboa - UL,
Av. Rovisco Pais 1, 1049-001 Lisboa, Portugal}
\affiliation[c]{Department of Physics, Aristotle University of Thessaloniki, 
54124 Thessaloniki, Greece}

\emailAdd{christos.charmousis@ijclab.in2p3.fr}
\emailAdd{antoine.lehebel@tecnico.ulisboa.pt}
\emailAdd{esmyrnio@auth.gr}
\emailAdd{niksterg@auth.gr}

\abstract{We study the properties of compact objects in a particular 4D Horndeski theory originating from higher dimensional  Einstein-Gauss-Bonnet gravity. Remarkably, an exact vacuum solution is known. This compact object  differs from general relativity mostly in the strong field regime. We discuss some properties of black holes in this framework and investigate in detail the properties of neutron stars, both static and in slow rotation. We find that for relatively modest deviations from general relativity, the secondary object in GW190814 is compatible with being a slowly-rotating neutron star, without resorting to very stiff or exotic equations of state. Remarkably, the equilibrium sequence of neutron stars matches asymptotically to the black hole limit, completetly closing the mass gap between neutron stars and black holes of same radius, although the stability of equilibrium solutions has yet to be determined. As a consequence, there exists a universal endpoint for the neutron star sequence, independent of the equation of state. In light of our results and of current observational constraints, we discuss specific constraints on the coupling constant that parametrizes deviations from general relativity in this theory.}

\begin{document}
\maketitle

\section{Introduction}

In recent years, there is a plethora of observational data on compact objects (neutron stars and black holes) coming from gravitational-wave (GW) detectors, such as Advanced LIGO \cite{LIGOScientific:2014pky}, Advanced Virgo \cite{VIRGO:2014yos}, from networks of radiotelescopes, such as the Event Horizon Telescope \cite{ EventHorizonTelescope:2020qrl} and from X-ray observations, such as the NICER mission \cite{Raaijmakers:2019dks}, to name a few.  Within current observational uncertainties, all data are compatible with general relativity (GR), but there is room for an alternative theory of gravity to emerge as a preferred choice in future observations with smaller uncertainties. Apart from direct tests of GR, there are also outstanding questions to be answered, such as the high-density equation of state (EOS), the maximum mass of neutron stars, etc. The recent observation of the compact object merger GW190814 \cite{LIGOScientific:2020zkf} indicates that the lighter member of the binary has a mass of $2.59_{-0.09}^{+0.08}~M_\odot$, which places it in the current observational mass gap between neutron stars and black holes and which could be explained (within GR) only as a neutron star with an unexpectedly stiff (or exotic) EOS, a neutron star with an unexpectedly rapid rotation, or a black hole with an unexpectedly small mass, see \cite{Huang:2020cab,Bombaci:2020vgw,Roupas:2020nua,Zhou:2020xan,Most:2020bba,Tan:2020ics,Vattis:2020iuz,Zhang:2020zsc,Fattoyev:2020cws,Tsokaros:2020hli,Tews:2020ylw,Dexheimer:2020rlp,Godzieba:2020tjn,Kanakis-Pegios:2020kzp,Nathanail:2021tay,Roupas:2020jyv,2021MNRAS.505.1600B}. Alternatively, the high mass of the secondary component in GW190814 could be explained in the framework of an alternative theory of gravity, see e.g. \cite{Nunes:2020cuz,Astashenok:2020qds,Astashenok:2021peo}.

So far, two GW events have been identified as binary neutron star (BNS) mergers, GW170817 \citep{2017PhRvL.119p1101A} and GW190425 \citep{2020ApJ...892L...3A}, and more are expected in the next years \citep{Aasi:2013wya}. Detection of GW from the inspiral phase of GW170817, in combination with observations of its electromagnetic counterpart \citep{GBM:2017lvd,Monitor:2017mdv,Goldstein_etal_2017}, have produced new constraints on the dimensionless tidal deformability of neutron stars and thus on their EOS; see \citep{Bauswein_etal_2017,Abbott:2018exr,2020NatAs...4..625C,2020PhRvD.101l3007L,2020Sci...370.1450D,2021arXiv210101201B} 
and references therein, as well as \citep{2020GReGr..52..109C,2021GReGr..53...27D} for recent reviews. These EOS constraints are expected to improve by combining a larger number of detections in the near future \cite{DelPozzo2013,2015PhRvD..92j4008C,2015PhRvD..91d3002L,2019PhRvD.100j3009H,2020PhRvD.101d4019C}. Although the sensitivity of the Advanced LIGO and Advanced Virgo detectors was not sufficient to detect the post-merger phase in GW170817  \citep{2017PhRvL.119p1101A,2017ApJ...851L..16A}, such a detection is likely to be achieved in the future, with upgraded \citep{KAGRA:2020npa}, with dedicated high-frequency \citep{2021PhRvD.103b2002G,2021CmPhy...4...27P} or with third-generation \citep{LIGOScientific:2016wof,Maggiore:2019uih} detectors. The observation of GW in the post-merger phase of a BNS merger would then offer another opportunity to probe the high-density EOS, see \citep{1992ApJ...401..226R,2005PhRvL..94t1101S,PhysRevLett.108.011101,2014PhRvD..90f2004C,PhysRevD.93.124051,2019JPhG...46k3002B,2019PhRvD.100j4029B,PhysRevD.100.044047,2020PhRvD.101h4039V,2020PhRvD.102d3011E,2020IJMPD..2941015F} and references therein.

Given the inflow of present and future observational data, it is important to study compact objects in specific and well-defined alternative theories of gravity. This in order to provide measurable deviations to the well-known GR solutions of compact objects, while additionally providing signatures going beyond GR. In this paper we will study a specific Horndeski \cite{Horndeski:1974wa} scalar-tensor model, which presents a number of attractive properties/symmetries and ---~most probably as a result of these~--- interesting analytic solutions. The action that we consider is 
\begin{equation}
S=\dfrac{1}{2\kappa}\displaystyle\int{\mathrm{d}^4x\sqrt{-g}\left\{R+\alpha\left[\phi\mathcal{G}+4G_{\mu\nu}\nabla^\mu\phi\nabla^\nu\phi-4(\nabla\phi)^2\Box\phi+2(\nabla\phi)^4\right]\right\}}+S_\mathrm{m},
\label{eq:action}
\end{equation}
where $\kappa=8\pi G/c^4$, $\mathcal{G}=R^2-4R_{\mu\nu}R^{\mu\nu}+R_{\mu\nu\rho\sigma}R^{\mu\nu\rho\sigma}$ is the Gauss-Bonnet scalar and $S_\mathrm{m}$ is the matter Lagrangian. We take the scalar to be dimensionless, hence $\alpha$ has the dimension of a length squared. Within this theory, we will first study and extend the black hole solution recently found in \cite{Lu:2020iav}. This solution is observationally very close to its GR counterpart \cite{Clifton:2020xhc}, but has intriguingly different structure and theoretical properties.  We will then construct models of neutron stars, using two different numerical methods. We will see that this theory quite naturally allows for more massive static neutron stars than GR while not being restrictive on the equations of state; the mass of the secondary component in GW190814 could be explained without resorting to very stiff or exotic EOS.

The theory \eqref{eq:action} has interesting properties which are quite different from those encountered in previously studied Horndeski models. As we mentioned, there exists an exact black hole solution \cite{Lu:2020iav} (see also the complementary analyses in \cite{Hennigar:2020lsl,Fernandes:2021ysi}) which is genuinely different from the Schwarzschild solution, while it is close enough to GR in order to pass observational tests coming from local or weak gravity constraints \cite{Clifton:2020xhc} (for reasonable values of the relevant coupling constant $\alpha$). The solution stands out from the numerous known analytic solutions of Horndeski theory, first because it is intrinsically different from the so-called stealth solutions, and from the scalarization phenomenon. Quite generically, Horndeski theory has been found to admit stealth black hole solutions, i.e. black holes with an identical metric to that of GR and a non-trivial scalar field. These stealth solutions were first found for a particular Horndeski model, assuming spherical symmetry \cite{Babichev:2013cya}, and then extended to generic Horndeski \cite{Kobayashi:2014eva} and beyond \cite{Motohashi:2019sen} theories. Within this family, stealth solutions have been recently extended to stationary spacetimes \cite{Charmousis:2019vnf}. 
This was achieved by realising that the scalar accompanying the stealth solutions is related to the integrable geodesics of Kerr spacetime \cite{Carter:1968ks}. Interestingly, stationary stealth solutions provide seed metrics in order to construct non-stealth stationary black holes with many surprising properties \cite{Anson:2020trg,BenAchour:2020fgy,Anson:2021yli}. 
In stark difference here, the black hole solution is not stealth, as the metric also differs from GR and the scalar field has no immediate relation to the black hole geodesics. We also point out that the theory \eqref{eq:action} does not have scalarization properties \cite{Damour:1993hw,Silva:2017uqg,Doneva:2017bvd}. In other words, 
GR solutions with a trivial scalar field are not solutions of the theory (\ref{eq:action}). Therefore, there cannot be a   transition from a GR branch to a scalarized branch of solutions.

The second outstanding property of the theory under consideration is that it is not parity symmetric with respect to the scalar field ($\phi\to-\phi$). Certain terms have even parity symmetry and others odd parity symmetry, according to the number of scalar derivatives. In terms of Horndeski theory, all terms are present in the action, $G_2$, $G_3$, $G_4$ as well as $G_5$ with the parametrization of Ref.~\cite{Kobayashi:2011nu}. Note that the theory \eqref{eq:action} preserves shift symmetry ($\phi\to\phi+C$ where $C$ is a constant). Analytic solutions, notably stealth solutions, are a common feature of parity and shift-symmetric models (see e.g. \cite{Babichev:2016rlq,Lehebel:2018zga} for reviews). However, when odd terms are present, only numerical solutions were previously available, for instance in the context of Einstein-scalar-Gauss-Bonnet gravity \cite{Kanti:1995vq,Campbell:1991kz,Sotiriou:2013qea}, or within cubic galileon models \cite{Babichev:2016fbg,VanAelst:2019kku,Emond:2019myx}. This is probably the reason why the solution we will study has genuinely different properties compared to parity preserving theories.

Finally, from a theoretical perspective, the theory \eqref{eq:action} has been constructed advocating differing symmetry properties: regularization via conformal symmetry and topology \cite{Hennigar:2020lsl,Fernandes:2020nbq,Fernandes:2021dsb} or, a more standard, but careful limiting process \cite{Lu:2020iav} of Kaluza-Klein compactification \cite{VanAcoleyen:2011mj,Charmousis:2012dw} of higher-dimensional metric Lovelock theories \cite{Lovelock:1971yv,Charmousis:2014mia}.  
This makes the model attractive theoretically, as it may be related to ultra-violet corrections of GR, such as those proposed by classical string theory effective actions (see for example \cite{Fradkin:1985fq,Gross:1986iv,Metsaev:1987zx}). 

Previous works have studied neutron stars in the context of Horndeski theory and its extensions, starting with the original Damour-Esposito-Farèse scalarization models \cite{Damour:1993hw}. The cubic \cite{Ogawa:2019gjc} and quartic \cite{Cisterna:2015yla,Cisterna:2016vdx} terms have been investigated in this respect. The quintic sector has also received a lot of attention, especially in the form of a coupling to the Gauss-Bonnet scalar, in the context of scalarization \cite{Doneva:2017bvd,Ventagli:2021ubn}. Interesting effects have been pointed out in the extension of Horndeski theory known as degenerate higher-order scalar-tensor (DHOST) theories. In particular, the Vainshtein screening mechanism breaks down inside matter \cite{Kobayashi:2014ida}, which was illustrated with neutron star configurations in \cite{Babichev:2016jom,Sakstein:2016oel}. Finally, neutron star solutions have been constructed in a cosmologically interesting subclass of DHOST theories \cite{Chagoya:2018lmv,Kobayashi:2018xvr}.
An important difference with the models considered in Refs.~\cite{Chagoya:2018lmv,Kobayashi:2018xvr}, is that, here, the metric and its first derivative will be continuous at the surface of the star.

The structure of the paper is as follows: In Sec.~ \ref{vacuum}, we start by analysing the properties of the vacuum solution found in \cite{Lu:2020iav} and extend it for a time-dependent scalar. We give a simple argument disqualifying negative coupling constant $\alpha$ (see also \cite{Clifton:2020xhc} for a cosmological argument). Then, in Sec.~\ref{sec:statNS}, using two distinct  numerical methods, we find neutron star solutions and compare their properties to their GR counterparts. Interestingly, within the constraints imposed on $\alpha$, one can attain with relative ease large mass static neutron star solutions with little or no constraint on common EOS.  In Sec.~\ref{rotating}, we discuss the case of slowly rotating compact objects, and we conclude in Sec.~\ref{discussion} with a discussion.

\section{Vacuum solution}
\label{vacuum}

Remarkably, the theory based on the action (\ref{eq:action}) possesses an exact vacuum solution \cite{Lu:2020iav,Hennigar:2020lsl,Fernandes:2020nbq}. Consider a spherically symmetric and static spacetime, with line element
\begin{equation}
\mathrm{d}s^2=-h(r)\mathrm{d}t^2+\dfrac{\mathrm{d}r^2}{f(r)}+r^2\mathrm{d}\Omega^2 ,
\label{eq:ansatz}
\end{equation}
where $h(r)$ and $f(r)$ are metric functions. The scalar field is also dependent only on the radial coordinate, $\phi=\phi (r)$. The field equations associated with \eqref{eq:action} and \eqref{eq:ansatz} are given in Appendix \ref{sec:fieldeqs}. The action \eqref{eq:action} with its particular couplings is quite special, as one combination of the field equations yields a pure geometric constraint, namely $2R+\alpha \mathcal{G}=0$, which can be written
\begin{equation}
\sqrt{\frac{f}{h}}\left[2 h' \sqrt{\frac{f}{h}}(r^2+2 \alpha (1-f)) \right]'+ \left[4 r (f-1) \right]'=0.
\end{equation}
The field equations admit the following analytic solution:
\begin{align}
h(r)&=f(r)=1+\dfrac{r^2}{2 \alpha }\left(1-\sqrt{1+ \dfrac{8 \alpha  M} {r^3}}\:\right),
\label{eq:vacf}
\\
\phi(r)&=\displaystyle\int{\mathrm{d}r\dfrac{\sqrt{f}- 1}{r\sqrt{f}}},
\label{eq:vacphi}
\end{align}
where $M$ is a free parameter. We will see that we can interpret $M$ as the Arnowitt-Deser-Misner (ADM) mass of the solution,\footnote{If we choose $M$ to have the dimension of a mass, we should replace $M$ by $MG/c^2$ in Eq.~\eqref{eq:vacf}. However, we will work in units where $c=1$ and $G=1$ throughout the paper, and restore them if needed.} hence we will restrict to positive $M$. There actually exist other branches of solutions, but this one is the only asymptotically flat solution with $\phi$ going to 0 at spatial infinity (or rather to a constant, since the action possesses a global shift symmetry for the scalar), and free of naked singularities. Reference \cite{Fernandes:2021ysi} has shown numerically that another branch of solutions, characterized by $\phi'=h'/(2h)$ and a priori compatible with flat asymptotics, always contains a naked singularity.

\subsection{Analytic properties}

 The above solution describes a black hole with an event horizon at
\begin{equation}
r_\mathrm{h}=M+\sqrt{M^2 -\alpha}
\label{eq:rh}
\end{equation}
if $\alpha<0$, or if $\alpha>0$ and $M>M_\mathrm{min}$ with
\begin{equation}
M_\mathrm{min}=\sqrt{\alpha}.
\label{eq:Mmin}
\end{equation}
If $\alpha>0$ and $M<M_\mathrm{min}$ on the other hand, $f$ vanishes nowhere. Hence, the solution possesses no horizon. At first sight, the metric looks regular in this case. $f$ has a minimum at 
\begin{equation}
r_\mathrm{min}=(M\alpha)^{1/3},
\end{equation}
and goes to 1 as $r$ approaches 0. However, the metric is actually singular close to $r=0$, because it is not locally homeomorphic to a de Sitter spacetime there. This can be seen by computing the Ricci scalar, which diverges:
\begin{equation}
R(r)\underset{r\rightarrow 0}{\sim}\dfrac{15}{4}\sqrt{\dfrac{2M}{\alpha r^3}}.
\label{eq:Rdiv}
\end{equation} 
Note that, in the case where the solution possesses a horizon, the scalar field \eqref{eq:vacphi} is ill-defined inside the black hole. The interior solution must therefore be different. When approaching the horizon, $\phi$ remains finite while $\phi'$ diverges. However, $\phi'$ is not a coordinate invariant quantity. The corresponding scalar quantity is $(\nabla\phi)^2$, which is finite at the horizon. On the other hand, scalar quantities built with higher than first order derivatives of the scalar, such as $\Box\phi$, diverge at the horizon. It is therefore not entirely obvious whether the solution can be continued analytically inside the horizon; in the next section, we provide a way to make the solution regular everywhere inside the black hole. When $\alpha<0$, $f$ could a priori become singular for $r<2(-\alpha M)^{1/3}$, where the quantity under the square root in Eq.~\eqref{eq:vacf} becomes negative. However, this always happens when $r<r_\mathrm{h}$, and we can no longer trust the solution in this region, as we just explained. At the other end of the radial coordinate span,
\begin{align}
\label{label}
f(r)&\underset{r\rightarrow +\infty}{=}1-\frac{2 M}{r}+\frac{4 \alpha  M^2}{r^4}+\mathcal{O}(r^{-5}),
\\
\phi(r)&\underset{r\rightarrow +\infty}{=}\dfrac{M}{r}+\mathcal{O}(r^{-2}),
\end{align}
which justifies our interpretation of $M$ as the ADM mass of the solution.

\subsection{Time-dependent solution}

One way to circumvent the problem we encounter with the regularity of the scalar within the horizon is to extend the solution to a time-dependent scalar. It is well known that, for models that possess shift symmetry $\phi\to\phi+C$ (where $C$ is some arbitrary constant), one can consistently allow the scalar field to depend linearly on time, while keeping a static metric \cite{Babichev:2015rva}. This possibility is particularly interesting in light of cosmology, where the asymptotics naturally dictate that the scalar field depends on time. It is remarkable that, considering a linearly time-dependent ansatz for the scalar field, the metric functions given in Eq.~\eqref{eq:vacf} remain solutions to the field equations. Only the scalar field is modified, according to
\begin{equation}
\phi=qt+\displaystyle\int{\mathrm{d}r\:\dfrac{f-\sqrt{f+q^2 r^2}}{r f}},
\label{eq:phitdep}
\end{equation}
where $q$ is an arbitrary constant. It is quite intriguing that the metric is unaffected by this time dependence.{\footnote{See \cite{Babichev:2017guv} for a counter-example, where time dependence regularises the scalar in a similar way, but alters the metric and spoils its asymptotics.}} This persists even for the neutron star solutions and the slowly rotating solutions that we present in Secs.~\ref{sec:statNS} and \ref{rotating}. Indeed, in the case of neutron stars, the exterior solution is identical to the black hole one. In the interior, it is easy to check that Eq.~\eqref{eq:phitdep} is still a solution (it solves the $(tr)$ component of the metric equations). Upon plugging Eq.~\eqref{eq:phitdep} in the other metric equations, these become identical to the static case, $q=0$. Hence the metric and matter profile do not depend on the choice of $q$, only the scalar field does. This is a very distinctive property of the theory \eqref{eq:action}; for instance, the stealth solutions presented in \cite{Cisterna:2015yla} possess the same $q$-independence in the vacuum, but not when matter is present. Remarkably, in the black hole case, this time dependence, by making the term under the square root in Eq.~\eqref{eq:phitdep} strictly positive, makes the solution regular when crossing the horizon. 

In the case where $\alpha>0$, it is interesting to note that $f$, as defined in Eq.~\eqref{eq:vacf}, is such that, at any point, $0<f(r)<1$, with $f(0)=1$. Since nothing fixes the value of $q$, it is always possible to make the quantity under the square root of Eq.~\eqref{eq:phitdep} positive everywhere. Therefore, for $\alpha>0$, the black hole solution possesses an inner Cauchy horizon, at 
\begin{equation}
    r_-=M-\sqrt{M^2-\alpha}.
\end{equation}
The black hole becomes extremal, much like a Reissner-Nordstr\"om or Kerr black hole, at $M=M_\text{min}$, given in Eq.~\eqref{eq:Mmin}. Inside this inner horizon, the black hole possesses a singularity, characterised by the continuity of the metric at $r=0$, but still diverging curvature invariants, as shown by Eq.~\eqref{eq:Rdiv}. It is therefore more regular than the Schwarzschild solution at the origin, but yet not smooth enough to be a regular black hole (see for example \cite{Babichev:2020qpr,Baake:2021jzv} in the context of DHOST theories). This is clearly due to the higher order terms appearing in the action, which counter the divergence of the curvature as it occurs in GR.  For $\alpha<0$, and if we make $q$ large enough, we reach a singularity at $r_\text{sing}=2(-\alpha M)^{1/3}$. This singularity is always hidden behind the event horizon, but there is no inner horizon in this case.

Finally, let us remark that, when $q\neq0$, the second branch of solutions of the theory \eqref{eq:action}, with $\phi'=h'/(2h)$, does not admit asymptotically flat solutions. Therefore, the solution \eqref{eq:vacf}-\eqref{eq:phitdep} is the unique asymptotically flat and static vacuum solution of the theory \eqref{eq:action} when $\phi$ depends linearly on time.

\subsection{Consistency of the vacuum solution for small objects}
\label{sec:smallobj}

The weak field tests of the model \eqref{eq:action} have already been thoroughly examined in Ref.~\cite{Clifton:2020xhc}. We will not insist on these, but just discuss the consistency of the relative positions of various radii. This actually provides a drastic way to exclude negative values of $\alpha$. Indeed, the metric \eqref{eq:vacf} must be able to describe the exterior of common astrophysical objects, or simply objects that surround us. This can be consistently achieved only if the size $R$ of these objects is greater than $r_\mathrm{h}$ and $r_\mathrm{min}$. Indeed, if $R<r_\mathrm{h}$, the object would be shielded by a horizon. If $R<r_\mathrm{min}$, the issue comes from the matching between the exterior and the interior metrics. Let us consider a perfect fluid for simplicity. It obeys the conservation equation $\nabla_\mu T^{\mu\nu}=0$, which, for a spherically symmetric and static background, is given by Eq.~\eqref{eq:divT}. At the surface of the object, the pressure vanishes by continuity (otherwise pressure forces would blow off the external layers). Assuming that $P>0$ and $\rho>0$, the pressure can only increase when entering the object, or in other words $P'(R)<0$. Equation \eqref{eq:divT} in turn implies that $h'(R)>0$. However, if $R<r_\mathrm{min}$, the exterior solution is such that $h'(R)<0$ ($r_\mathrm{min}$ being the unique minimum of the exterior solution). Thus, the interior solution can only be matched to the exterior one if $R>r_\mathrm{min}$.

Let us start by considering $\alpha>0$. Equation \eqref{eq:rh} tells us that, if $r_\mathrm{h}$ exists, it is always smaller than $r_\mathrm{S}$, the Schwarzschild radius of the object under consideration (i.e., the size of a black hole in GR, with the same mass as this object). For objects that generate weak curvature, we thus clearly have $R>r_\mathrm{h}$. Now, in order to compare $R$ and $r_\mathrm{min}$, let us consider an object in the weak field regime, with average mass density $\bar\rho$. It approximately fulfills,
\begin{equation}
\dfrac43\pi R^3 \bar\rho\simeq M.
\end{equation}
Hence, requiring that $R>r_\mathrm{min}$ amounts to
\begin{equation}
\label{eq:earthconstraint}
\alpha\lesssim\dfrac{3c^2}{4\pi G\bar\rho}=5.8\times10^{22}~\mathrm{m}^2 \left(\dfrac{\bar\rho}{\bar\rho_\mathrm{E}}\right)^{-1},
\end{equation}
where $\bar\rho_\mathrm{E}=5.51$~g/cm$^3$ is the average density of the Earth. This leaves a lot of room for positive $\alpha$ solutions.

On the other hand, for $\alpha<0$, the behavior of $r_\mathrm{h}$ at small $M$ is rather problematic. Indeed, consider for instance an atomic nucleus of size $R\simeq10^{-15}$~m. Since we can probe atomic nuclei, they should not be shielded by a horizon, which translates to $r_\text{h}<R$, or equivalently to,
\begin{equation}
-\alpha<R(R-2M).
\label{eq:alphaneg}
\end{equation}
In the case of an atomic nucleus, restoring $G$ and $c$, the gravitational radius $M$ is of order $10^{-54}$~m. Hence, Eq.~\eqref{eq:alphaneg} gives a compelling bound on negative values of $\alpha$, $-\alpha\lesssim10^{-30}$~m$^2$, making the associated gravitational effects totally undetectable. For practical purposes, we can therefore exclude negative $\alpha$ in our analysis.

\subsection{Black hole minimum mass: an upper bound on $\alpha$}

\begin{figure}[ht]
\centering
\includegraphics[width=\textwidth]{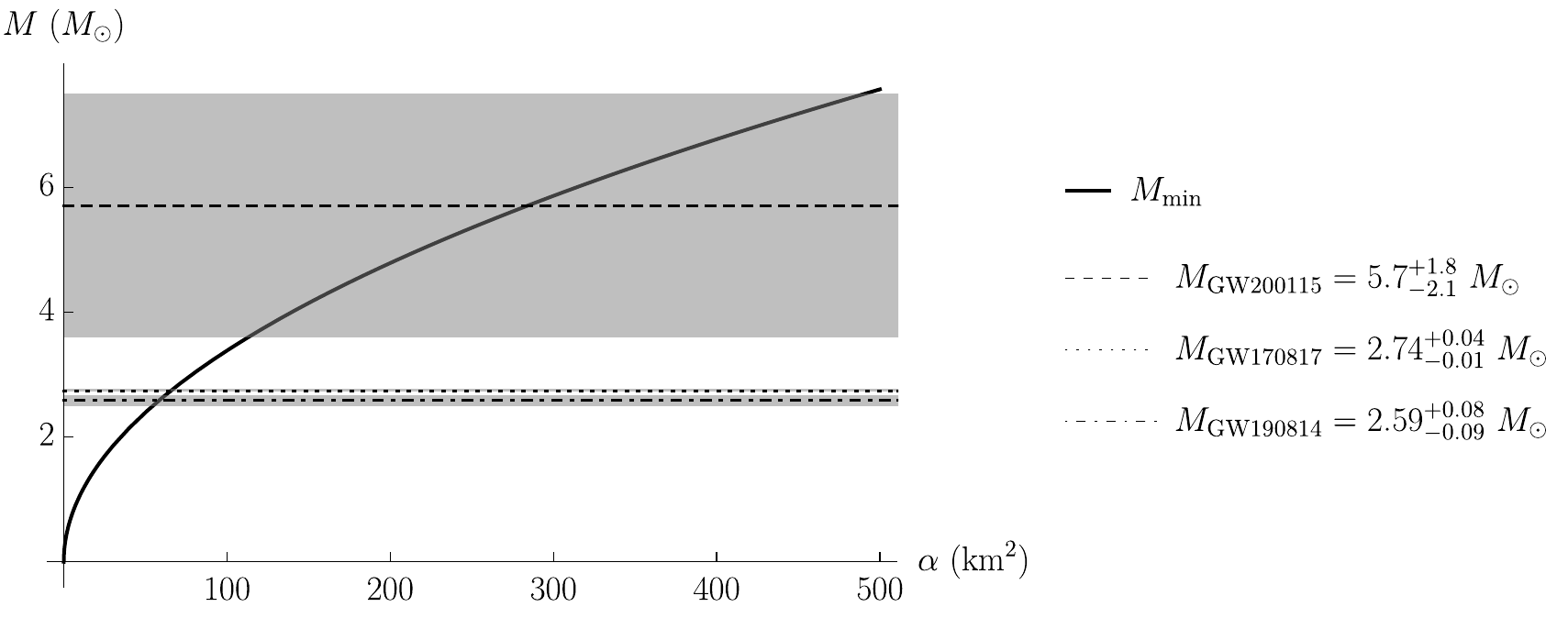}
\caption{Minimum possible black hole mass as a function of $\alpha$ (solid line). For positive values of $\alpha$, black holes cannot have a lighter mass than $M_\text{min}$, as given by Eq.~\eqref{eq:Mmin}. The dashed, dotted and dot-dashed lines correspond to the mass of some of the lightest black hole candidates, respectively observed in GW200115, GW170817 (maximum remnant black hole mass) and GW190814. The grey bands correspond to the $90\%$ credible interval for these observations (they are hardly visible in this plot for the latter two events). The black hole component of GW200115 can only exist in theories with $\alpha\lesssim285^{+207}_{-171}$~km$^2$. If the final object produced in GW170817 is a black hole, it requires that $\alpha\lesssim66$~km$^2$. Finally, if the lighter component of GW190814 is a black hole, it requires $\alpha\lesssim59$~km$^2$. All these constraints are estimates, relying on assuming a GR framework in the inspiral phase.}
\label{fig:MminBH}
\end{figure}

For positive values of $\alpha$, we already saw that black holes cannot have a lighter mass than $M_\text{min}$, given in Eq.~\eqref{eq:Mmin} (vacuum solutions below this mass exhibit a naked singularity that is not shielded by a horizon). However, the theory must be able to reproduce the lightest observed black holes. This sets an upper bound on $\alpha$. There exist several candidates for light black holes, from either GW or X-ray observations. Here, we will focus on the GW black hole detections, as the uncertainty in the mass determination is typically smaller than through X-ray observations \cite{Thompson:2018ycv}. In the GW200115 event, one of the components is identified as a black hole, with mass $M_\text{GW200115}=5.7^{+1.8}_{-2.1}~M_\odot$ at $90\%$ credible level \cite{LIGOScientific:2021qlt}. Note that, this mass estimate relies on the extraction of the chirp mass and the mass ratio in the inspiral phase, \textit{assuming GR}. A  fully  consistent treatment  would  require  re-analysing  the  inspiral in  the framework of the theory \eqref{eq:action}. We show in Fig.~\ref{fig:MminBH} how $M_\text{GW200115}$ compares with $M_\text{min}$ depending on $\alpha$.
To accommodate such a black hole, one needs
\begin{equation}
    \alpha\lesssim285^{+207}_{-171}~{\rm km}^2
\end{equation}
(90\% credible interval). There also exist lighter black hole candidates with a better precision on the mass measurement, but for which the exact nature of the object is less certain. Among these, if the remnant of the GW170817 BNS merger resulted in a delayed collapse to a black hole, its mass is smaller than the total mass,\footnote{In reality, it will be even lower, due to gravitational wave emission, either tensor or scalar (since both can be present in the model we study).} $M_\text{GW170817} < 2.74^{+0.04}_{-0.01}~M_\odot$ \cite{LIGOScientific:2017vwq}. This imposes
\begin{equation}
    \alpha\lesssim66~{\rm km}^2.
\end{equation}
Finally, the lightest component of the GW190814 event has a mass $M_\text{GW190814}=2.59^{+0.08}_{-0.09}~M_\odot$ \cite{LIGOScientific:2020zkf}. If this object is a black hole, this requires 
\begin{equation}
    \alpha\lesssim59~{\rm km}^2.
\end{equation}
We will see in the next section that a very similar bound can be placed on $\alpha$ using neutron star observations, rather than black holes. The advantage of the bound derived from black holes, is that it is completely independent of the EOS and hence of the tidal deformability of neutron stars.

\section{Static neutron stars constraints}
\label{sec:statNS}

We now move to the study of nonrotating (static) neutron star solutions based on the action \eqref{eq:action}. We model the interior of the neutron star as a perfect fluid with energy density $\epsilon$, pressure $P$ and 4-velocity $u^\mu$. Its stress-energy tensor is then given by
\begin{equation}
T_{\mu\nu}=(\epsilon+P)u_\mu u_\nu+Pg_{\mu\nu}.
\end{equation}
We will use several microphysical, cold (barotropic) equations of state $P=P(\epsilon)$, and implement them in the form of a high-order analytic fit, see \cite{2004A&A...428..191H} for the SLy EOS and  \cite{2011arXiv1108.2166G} for the AP4, MPA1 and MS2 EOS. These four EOS roughly span the allowed range of radii for typical neutron stars, as constrained by GW170817 \cite{Abbott:2018exr} and have a maximum mass of static models larger than the mass of pulsar PSR J0348+0432, which  was determined to have a mass of $2.01\pm0.04~M_\odot$ \cite{Antoniadis:2013pzd}. The fluid is at rest in the frame that we use, so that
\begin{align}
u_{\mu}&=(-\sqrt{h},0,0,0),
\\
T_{\mu\nu}&=\mathrm{Diag}(h\epsilon,P/f,r^2P,r^2\sin^2\theta P).
\end{align}
The scalar field equation is a conservation equation for the current $J^\mu$ of the scalar field, $\nabla_\mu J^\mu=0$ (see Appendix \ref{sec:fieldeqs}). In the geometry that we consider, only the radial component of the current is non-trivial. However, in order for  $J^\mu$ to be a solution of the conservation equation (especially at $r=0$), we must have $J^r=0$, i.e.
\begin{equation}
J^r=\dfrac{2 \alpha  }{\kappa r^2} \left[(r \phi'-1)^2 f-1\right] \left(2 \phi' h-h'\right)=0,
\end{equation}
even inside the star. For completeness let us note that the full scalar field equation can be integrated locally as $J^r r^2 \sqrt{f/h}=c$, where $c$ is a constant of integration  corresponding to scalar primary hair. However, if we do not set $c=0$, this is a solution to $\nabla_\mu J^\mu=c\:\delta(r)$ (with $\delta$ the Dirac distribution) rather than $\nabla_\mu J^\mu=0$. Hence $c=0$ in agreement with the no hair theorem for stars \cite{Lehebel:2017fag}. Furthermore, the generic solution for $q\neq0$ immediately imposes $J^r=0$ from the metric equations \cite{Babichev:2015rva}. At the end, by continuity with the exterior solution, $\phi$ is still determined by Eq.~\eqref{eq:vacphi} inside the star. This greatly simplifies the calculations, as we see that $\phi$ can be determined analytically rather than numerically.

To determine the metric, energy and pressure profiles, we need to integrate numerically the field equations. To this end, and for numerical consistency, we employed two very different numerical schemes. In the main text, we present one of these methods. It makes use of the Mathematica built-in solver NDSolve, and solves a system of first-order differential equations as an initial value problem, integrating from the center of the star to its surface, see Appendix \ref{sec:fieldeqs}. The second method, solves a system of elliptic equations as a boundary value problem iteratively in isotropic coordinates (a first integral of the hydrostatic equilibrium equation is included in the iteration procedure). This is based on the Komatsu-Eriguchi-Hachisu (KEH)/Cook-Shapiro-Teukolsky (CST) scheme \cite{10.1093/mnras/237.2.355,pub.1058503139} for rotating stars, see also \cite{friedman_stergioulas_2013}, and it is presented in detail in Appendix \ref{sec:altmethod}, where we also discuss the numerical agreement between results obtained with the two methods. It is the first time that the second method is implemented in the context of Horndeski theory. Beyond allowing a cross-check, it will also be useful for future studies of theories where the system of equations naturally comes with equations that are harder to cast in an explicit first-order form (typically non-shift symmetric models). It also converges faster for such theories.

\begin{figure}[ht]
\centering
\includegraphics[width=\textwidth]{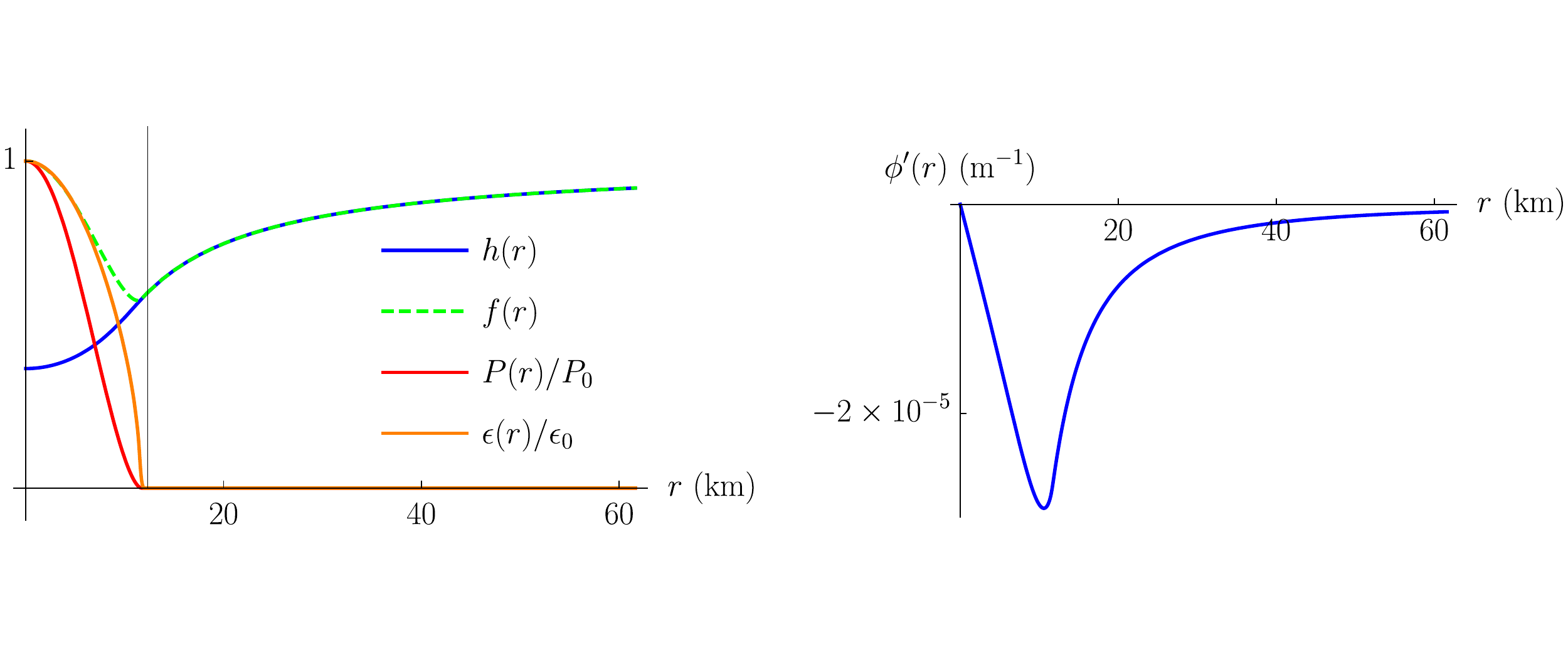}
\caption{Example of solution for $\alpha=10$~km$^2$, and a central density $\epsilon_0=10^{15}$~g/cm$^3$, using the SLy EOS. The left panel shows the metric functions $f(r)$ and $h(r)$, as well as the pressure $P(r)$ and energy density $\epsilon(r)$ distributions, normalized to their central values; the right panel shows the scalar profile.  The black vertical line in the left panel represents the position of the surface of the star. The star has a radius of $R=12.3$~km, and a mass of $1.73~M_\odot$.The numerical solution matches smoothly with the exterior exact solution \eqref{eq:vacf}-\eqref{eq:vacphi} at $r=R$.}
\label{fig:starexample}
\end{figure}

In the first method, the numerical solution is matched to the exact exterior solution, Eq.~\eqref{eq:vacf}, at the surface of the star (where the pressure vanishes).\footnote{In practice, we must choose a cutoff density where we stop the integration, see Appendix \ref{sec:fieldeqs} for details.} The matching with the exterior solution is performed simply by requiring continuity of the metric and its first derivatives at the surface.  Close to the origin, on the other hand, we determine the boundary conditions by a series expansion:
\begin{align}
f(r)&\underset{r\to0}= 1-\dfrac{\sqrt{12 \alpha \kappa  \epsilon_0+9}-3}{6 \alpha}\;r^2+\mathcal{O}(r^4),
\label{eq:fc}
\\
h(r)&\underset{r\to0}= h_0\left[1+\dfrac{-\alpha \kappa \epsilon_0+\sqrt{12 \alpha \kappa \epsilon_0+9}+3 \alpha \kappa  P_0-3}{2 \alpha \sqrt{12 \alpha \kappa \epsilon_0+9}}\;r^2\right]+\mathcal{O}(r^4),
\label{eq:hc}
\\
P(r)&\underset{r\to0}= P_0-\dfrac{(P_0+\epsilon_0) \left(-\alpha \kappa \epsilon_0+\sqrt{12 \alpha \kappa \epsilon_0+9}+3 \alpha \kappa  P_0-3\right)}{4 \alpha \sqrt{12 \alpha \kappa \epsilon_0+9}}\;r^2+\mathcal{O}(r^4),
\label{eq:Pc}
\end{align}
where $P_0$ is the central pressure, $\epsilon_0$ is the central energy density and $h_0$ is an arbitrary scaling constant, that is fixed by the matching with the exterior solution. More details are given in Appendix \ref{sec:fieldeqs}. In passing, the above equations imply that $\epsilon_0<-3/(4\alpha\kappa)$ for negative $\alpha$. As we saw in paragraph \ref{sec:smallobj}, there are reasons to disregard negative values of $\alpha$. However, we will also consider such values in this section, and constrain them independently. A typical solution is shown in Fig.~\ref{fig:starexample}.
We obtain different solutions when varying the central density $\epsilon_0$ and the value of $\alpha$. Typically, deviations with respect to GR are marginal for $|\alpha|\lesssim1$~km$^2$. 

\begin{figure}[ht]
\centering
\includegraphics[width=\textwidth]{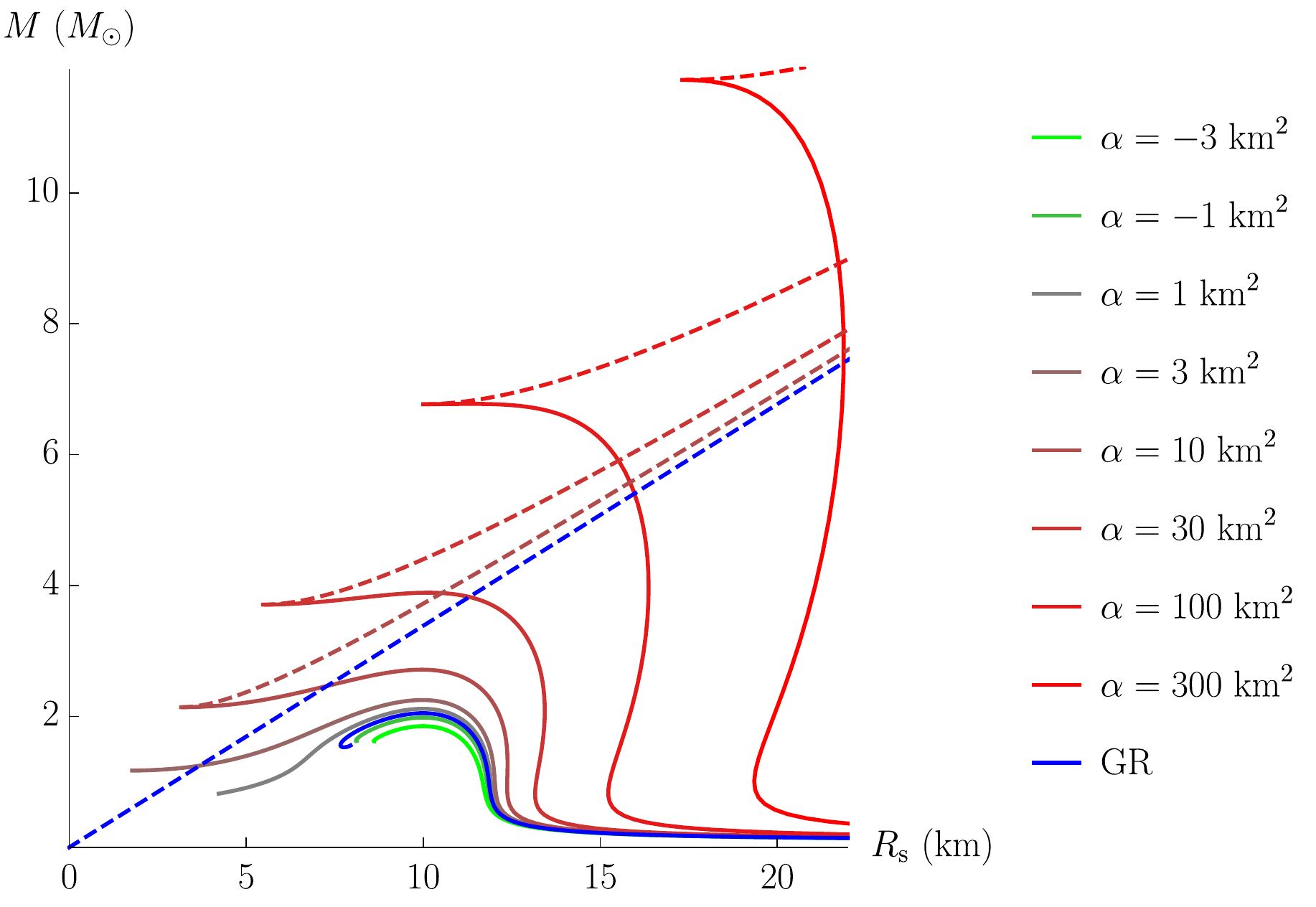}
\caption{Mass-radius relations for the SLy EOS and various values of $\alpha$. The plain blue line corresponds to GR solutions. The other plain lines correspond to nonzero values of $\alpha$; the color of these lines becomes closer to red (respectively green) as the value of $\alpha$ becomes more positive (respectively negative). The dashed lines correspond to the location of black hole solutions. The same color is used for a given value of $\alpha$, both for neutron star (plain lines) and black hole (dashed lines) solutions. We plotted only a selection of black hole lines in order to avoid overloading the figure. Generically, a positive $\alpha$ leads to larger masses and radii, while a negative $\alpha$ has the opposite effect. For non-zero $\alpha$, the neutron star existence lines asymptote black hole existence lines at high densities.}
\label{fig:MR}
\end{figure}

In Fig.~\ref{fig:MR}, we display the mass-radius relation of stars for the SLy EOS and various values of the coupling $\alpha$.
Similar plots can be obtained for the other EOS. Positive values of $\alpha$ increase both the mass and radius of neutron stars, with respect to GR. For values of $\alpha$  larger than about 10~${\rm km}^2$ and up to several tens ${\rm km}^2$, the radius for typical neutron stars remains within current observational radius constraints, while at the same time allowing for neutron star masses that are in the current observational mass gap between neutron stars and black holes, even for a relatively soft EOS, such as SLy. Therefore, the secondary component in GW190814 could have been a slowly rotating neutron star without resorting to extremely stiff or exotic EOS. In the same spirit, models with positive $\alpha$ can be used to alleviate the tension on the EOS that struggle to produce heavy enough neutron stars in GR. Such EOS include AP1-2, SQM1-3, PAL6 to name a few.

Very large positive values of $\alpha$ lead to extremely heavy neutron stars, that are unlikely to exist. Indeed, the most massive isolated neutron star (for which a precise mass measurement is available) is PSR J0348+0432, with a mass of $2.01\pm0.04~M_\odot$ \cite{Antoniadis:2013pzd}. If the lighter object in GW190814 was a neutron star, it would have a mass of $\sim 2.6~M_\odot$. Currently, there is no observational support for even higher neutron star masses. Positive values of $\alpha$ larger than $\sim 100~{\rm km}^2$ are also in strong tension with current constraints on neutron star radii. 

In GR, the turning-point theorem \cite{1988ApJ...325..722F} allows us to detect stability to collapse by inspection of the mass-radius relation for a sequence of non-rotating equilibrium models. In the alternative theory examined here, there is no analogous proof, as yet. Nevertheless, it is intriguing that for $\alpha\gtrsim30$~km$^2$, the turning point appears to the left of the $R=2M$ line. If the turning point is also indicative of stability (as is the case in GR) then this would mean that the theory we examine here allows for more compact stable neutron stars than a Schwarzschild black hole in GR. Of course, in this alternative theory, black holes are also modified and the black hole limit is displaced to higher values of compactness (dashed curves in  Fig.~\ref{fig:MR}).

\begin{figure}[ht]
\centering
\includegraphics[width=\textwidth]{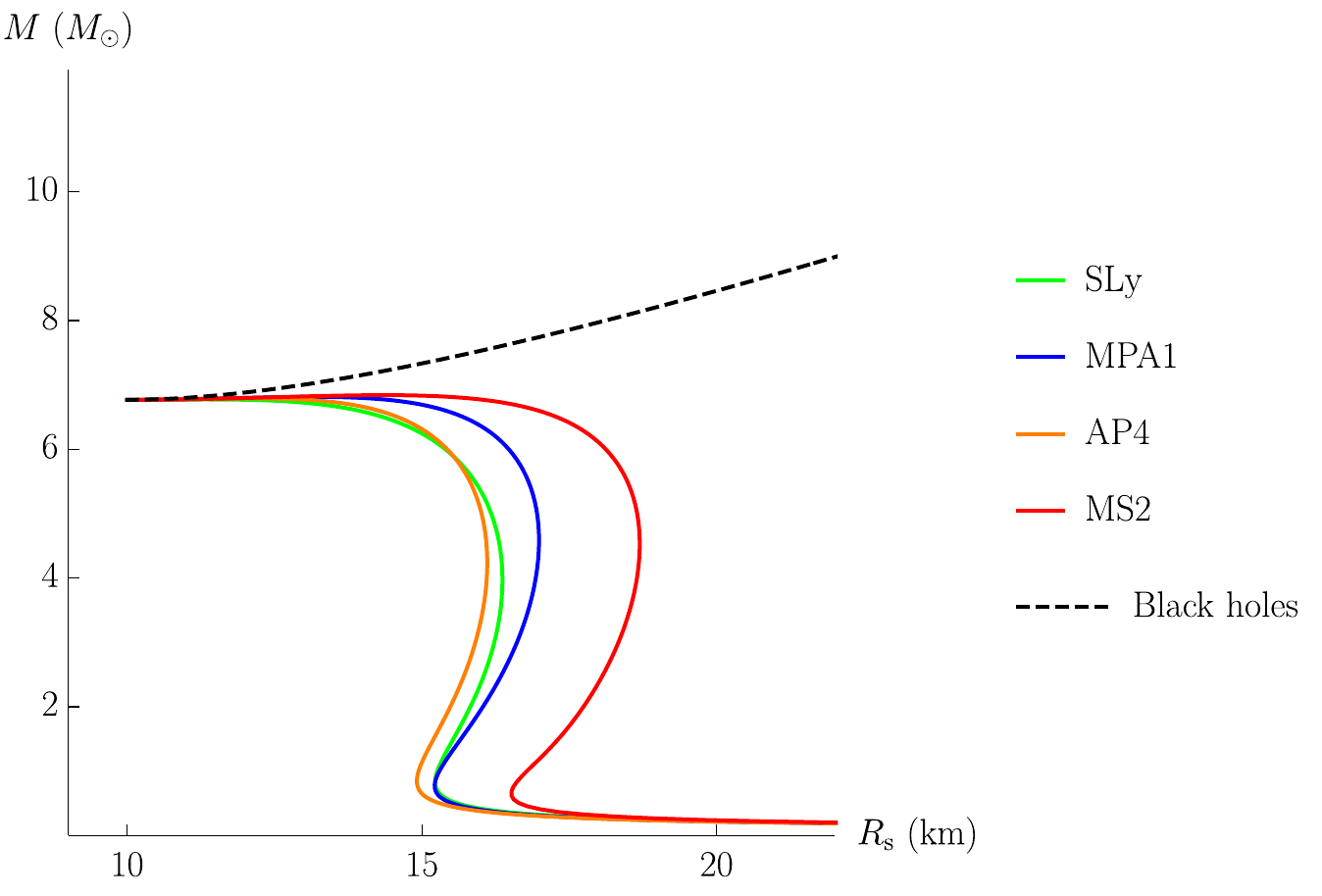}
\caption{Mass-radius relations for $\alpha=100$~km$^2$ and various EOS. The black dashed line corresponds to black hole solutions, while colored lines correspond to neutron star solutions. In spite of the variety of behaviors at lower densities, the neutron star sequences associated to different EOS universally converge towards the same endpoint, which is also the endpoint of the black hole sequence. This is a very unique feature of the theory under study.}
\label{fig:MREOS}
\end{figure}

There are even more drastic qualitative differences with GR. At high densities, the neutron star equilibrium sequence approaches the black hole limit asymptotically. Through detailed numerical investigations, we determined that the two sequences become arbitrarily close near the endpoint of the black hole existence curve. This behavior is specific to this alternative theory of gravity and does not occur in GR. It would be interesting to investigate the stability properties of such neutron star solutions. If they turn out to be stable, this would mean that there would be no mass gap between neutron stars and black holes of same radii. Remarkably, this also means that, for a fixed value of $\alpha$, there exists a universal endpoint for the neutron star sequence, which does not depend on the EOS. This behavior is illustrated in Fig.~\ref{fig:MREOS} for $\alpha=100$~km$^2$. It is a very distinctive feature of the theory \eqref{eq:action}, and we are not aware of other theories displaying this feature.

\begin{figure}[ht]
\centering
\includegraphics[width=\textwidth]{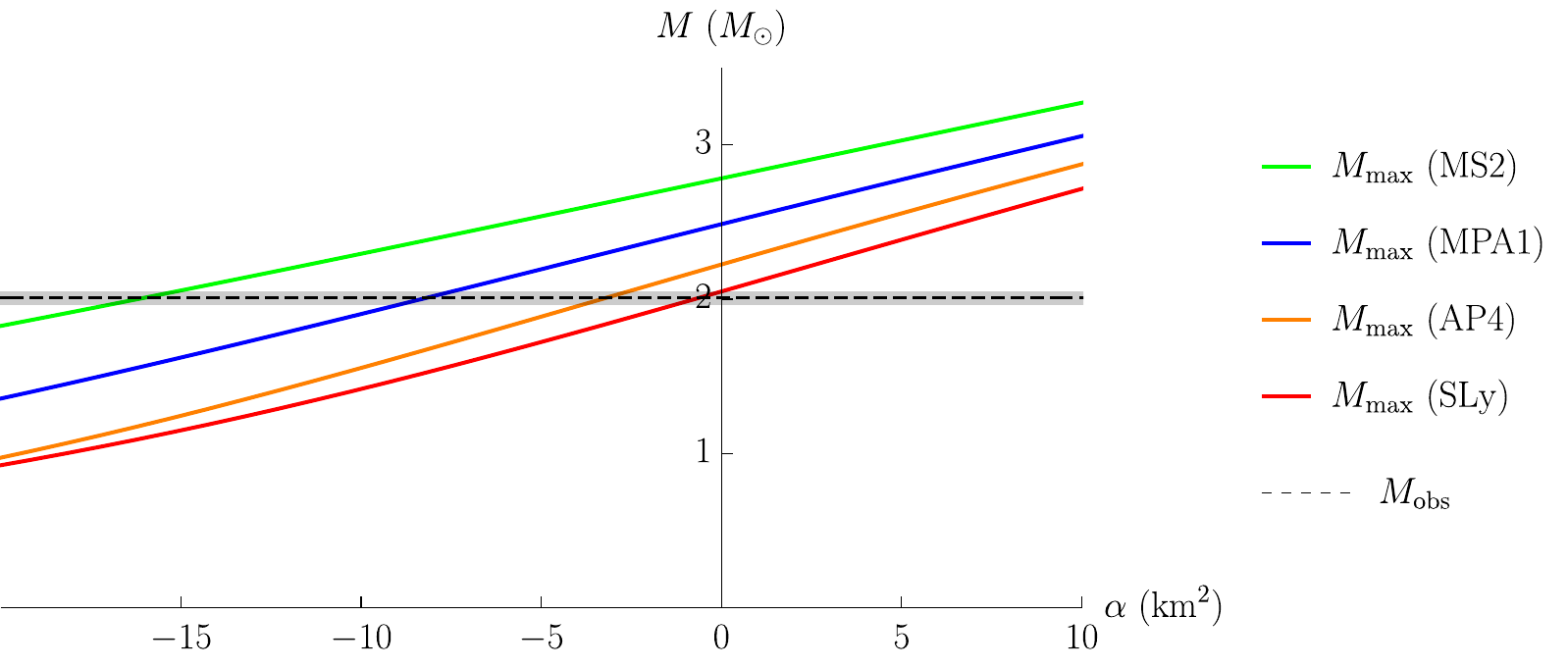}
\caption{Maximum static mass versus $\alpha$ for various EOS. The dashed line (with error bars) line is the mass of PSR J0348+0432, $M=2.01\pm0.04~M_\odot$. Requiring $M_{\rm max}>2.01~M_\odot$ constrains $\alpha$ to be larger than $\alpha_{\rm min}$, which is the intersection of the colored lines with the horizontal dashed line.}
\label{fig:Mmax}
\end{figure}

For smaller values of $\alpha$, we can derive a neutron star maximal mass. A satisfactory model should then be able to reproduce the mass of PSR J0348+0432, the heaviest pulsar with a precise mass measurement. In Fig.~\ref{fig:Mmax}, we plotted the maximum mass that can be reached for a neutron star at a given $\alpha$, for various EOS. 
Consider for instance the AP4 EOS. It is only for $\alpha\gtrsim-2.5$~km$^2$ that the associated curve is above the gray band, which corresponds to the mass of PSR J0348+0432. Therefore, for this EOS, $\alpha$ is bounded accordingly. For all EOS, this constraint gives a lower bound, $\alpha_\mathrm{min}$, on $\alpha$. These bounds are summarized in Table \ref{table:rangealpha}.
\begin{table}[ht]
\caption{Constraints on $\alpha$ coming from the requirement that static neutron stars must be allowed to have a mass larger than 2.01~$M_\odot$, for various EOS.}
\begin{center}
\begin{tabular}{l|c}
\hhline{==}
EOS&$\alpha_\mathrm{min}$ (km$^2$) \\
\hline
SLy & 0 \\

AP4& -2.5 \\

MPA1&-7.5 \\

MS2&-15.5 \\
\hline
\end{tabular}
\end{center}
\label{table:rangealpha}
\end{table}

\begin{figure}[ht]
\centering
\includegraphics[width=\textwidth]{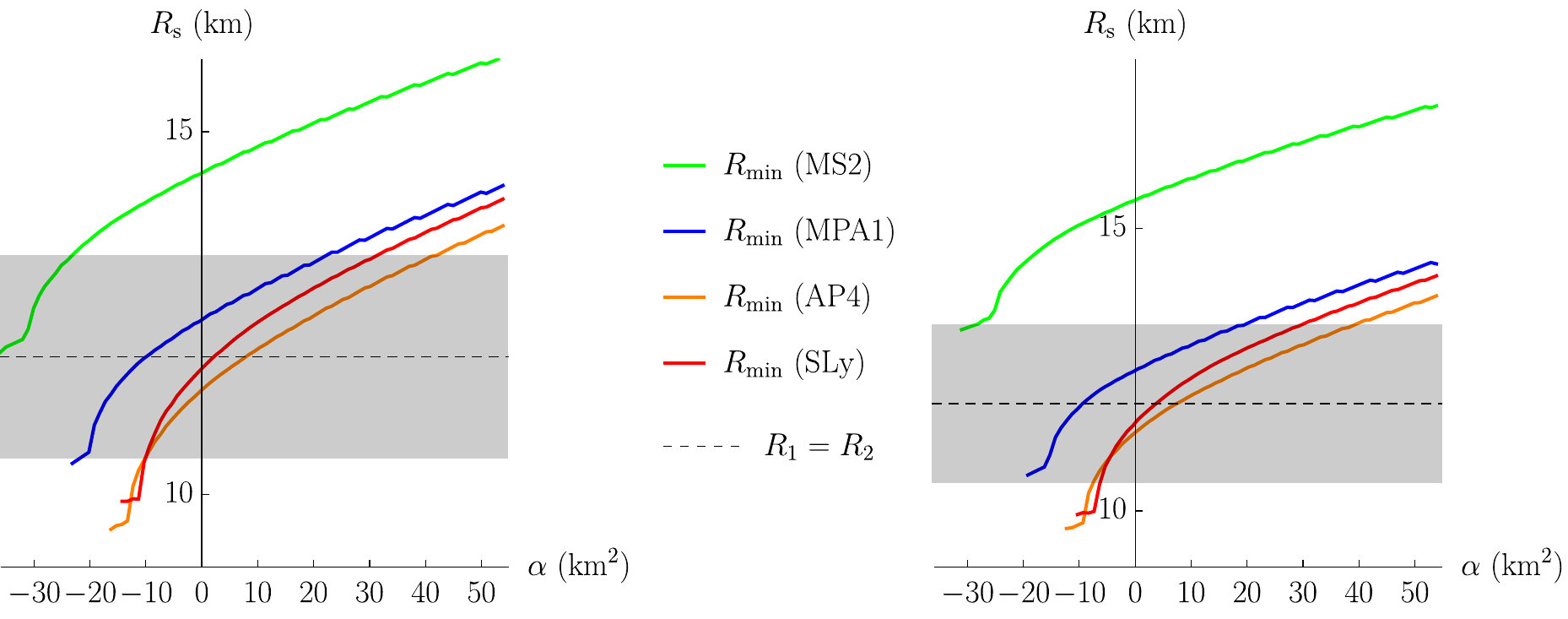}
\caption{Neutron star minimal radii versus $\alpha$ for various equations of state. The left (respectively right) panel shows the minimal radius for stars such that $1.17~M_\odot<M<1.36~M_\odot$ (respectively $1.36~M_\odot<M<1.6~M_\odot$). The color code is the same as in Fig.~\ref{fig:Mmax}. The dashed black line is the inferred radius of the components of GW170817 at 90$\%$ confidence level: $R_1=R_2=11.9\pm1.4$~km (the shaded region corresponds to the error bars). The minimal radius needs to be below $R_1=R_2=11.9\pm1.4$~km, which sets an upper bound on $\alpha$.}
\label{fig:Rmin}
\end{figure}

We can also use the mass-radius relations that we obtained to place an upper bound on $\alpha$, although this is more delicate. To this end, let us focus on the neutron star binary merger GW170817, which resulted in a radius constraint for the two neutron stars of $R_\mathrm{1}\simeq R_2=11.9\pm1.4$~km, with component masses $1.36~M_\odot< M_1<1.60~M_\odot$ and $1.17~M_\odot< M_1<1.36~M_\odot$ (low spin prior, at 90$\%$ confidence level) \cite{TheLIGOScientific:2017qsa,Abbott:2018exr}. As in the black hole case, a fully consistent treatment would require re-analysing the inspiral of GW170817 in the framework of the theory \eqref{eq:action}, and not in the framework of GR. We will use the above values of the masses and radii of the two components in the binary, in order to arrive at some first, {\it order of magnitude} constraints for the coupling constant $\alpha$. We therefore extract from the mass-radius relations the minimal radius in the ranges that are allowed for $M_1$ and $M_2$. This provides us a minimal radius $R_\text{min}$, for each value of $\alpha$ (and for each EOS). In Fig.~\ref{fig:Rmin}, we show, for each of the components, when $R_\text{min}$ is smaller than the estimated radius of $11.9\pm1.4$~km.
This constraint turns into upper bounds on $\alpha$, summarized in Table \ref{table:rangealpha2}.
\begin{table}[ht]
\caption{Estimated upper bound on $\alpha$, assuming the neutron star radii derived from GW170817 in a GR framework, for various EOS.}
\begin{center}
\begin{tabular}{l|c}
\hhline{==}
EOS & $\alpha_\mathrm{max}$ (km$^2$)\\
\hline
SLy &  30\\

AP4 & 39.5\\

MPA1 &20 \\

MS2 & -28 \\
\hline
\end{tabular}
\end{center}
\label{table:rangealpha2}
\end{table}
We stress again that, since the radius constraints from GW170817 were derived in the framework of GR, the bounds in Table \ref{table:rangealpha2} are not strict bounds, but should only be taken as order of magnitude estimates.

\section{Slowly rotating solutions}
\label{rotating}

In the previous section, we considered static neutron stars. We will now extend our analysis to slowly rotating solutions. Two different aspects of rotating solutions can be interesting for observations. First, the vacuum solution \eqref{eq:vacf} differs from a black hole in GR, leading to deviations for objects in geodetic motion. It is important to quantify such deviations. Second, the moment of inertia of neutron stars will also be different from GR. Although the moments of inertia of neutron stars have not yet been constrained, it will probably become possible in the coming years. In particular, thanks to precision timing of radio pulses, the moment of inertia of the $1.338~M_\odot$ primary component of the double pulsar PSR J0737-3039A \cite{Burgay:2003jj} is expected to be measured with an uncertainty of order $10\%$ \cite{2004Sci...303.1153L,2005ApJ...629..979L,Kramer:2009zza}. 

Here, we will treat rotation as a perturbation to the static solutions presented in the previous section, following the Hartle-Thorne formalism \cite{Hartle:1967he,Hartle:1968si}. It is estimated that most pulsars can be correctly described in this formalism. It was first developed in the framework of GR, but it can be straightforwardly adapted to scalar-tensor theories. At first order in the rotation parameter, the metric can be parametrized as
\begin{equation}
\mathrm{d}s^2=-h(r)\mathrm{d}t^2+\dfrac{\mathrm{d}r^2}{f(r)}+r^2(\mathrm{d}\theta^2+\sin^2\theta\mathrm{d}\varphi^2)-2\delta\omega(r)r^2\sin^2\theta\mathrm{d}t\mathrm{d}\varphi,
\label{eq:metricrot}
\end{equation}
while the scalar field is unchanged at this order. $\delta$ is the expansion parameter, and $\omega$ is determined by the field equations. It represents the speed at which an observer in geodetic motion rotates because of frame-dragging. The stress-energy tensor is also affected at first order in $\delta$, through the 4-velocity of the fluid:
\begin{equation}
u_\mu=\left(-\sqrt{h},0,0,\delta\dfrac{r^2}{\sqrt{h}}\sin^2\theta[\Omega-\omega(r)]\right),
\label{eq:speedrot}
\end{equation}
where $\Omega$ is the angular velocity of the star, which is assumed to rotate uniformly. Plugging Eqs.~\eqref{eq:metricrot} and \eqref{eq:speedrot} into the field equations, and expanding at first order in $\delta$, only the $(t\varphi)$ component of the metric equations is left. It is a linear, second order, ordinary differential equation in $\varpi\equiv\Omega-\omega$:
\begin{equation}
\begin{split}
0&=2 r f \varpi '' \left(2 \alpha -2 \alpha  f+ r^2\right)^2 +\varpi '\left\{-20 \alpha  f^2 \left(2 \alpha + r^2\right)-r^4 \left(2 \alpha + r^2\right) \kappa(P+\epsilon)\right.
\\
&\quad\left.+20 \alpha ^2 f^3+4 f \left[2\left(5 \alpha ^2+2 r^4+5 \alpha r^2\right)+\kappa\alpha  r^4 (P+3 \epsilon)\right]\right\}
\\
&\quad-4 r^3 \varpi \kappa(P+\epsilon) \left(2 \alpha -2\alpha  f+ r^2\right).
\end{split}
\label{eq:tphi}
\end{equation}
Remarkably, this equation can be solved exactly for the vacuum background, Eq.~\eqref{eq:vacf}. The solution is
\begin{equation}
\omega(r)= -\dfrac{J}{2 \alpha  M}\left(1-\sqrt{1+\dfrac{8 \alpha  M}{r^3}}\right)
\label{eq:omegavac}
\end{equation}
where $J$ is free, and should be interpreted as the total angular momentum of the star. Indeed, expanding Eq.~\eqref{eq:omegavac} at large $r$, we obtain
\begin{equation}
\omega(r)\underset{r\to\infty}{=} \dfrac{2 J}{r^3}-\dfrac{4\alpha  J M}{r^6}+\mathcal{O}(r^{-9}).
\label{eq:omegavac1}
\end{equation}
At leading order, the solution is thus identical to GR. Corrections are suppressed by a factor of $r^3$, hence they are extremely weak, and geodetic motion around a rotating star (or a black hole) will be almost indistinguishable from GR. We note in passing that Eq.~\eqref{eq:tphi}, and hence Eq.~\eqref{eq:omegavac}, remain valid even for  $q\neq 0$, the scalar field being unchanged (\ref{eq:phitdep}).

\begin{figure}[ht]
\centering
\includegraphics[width=\textwidth]{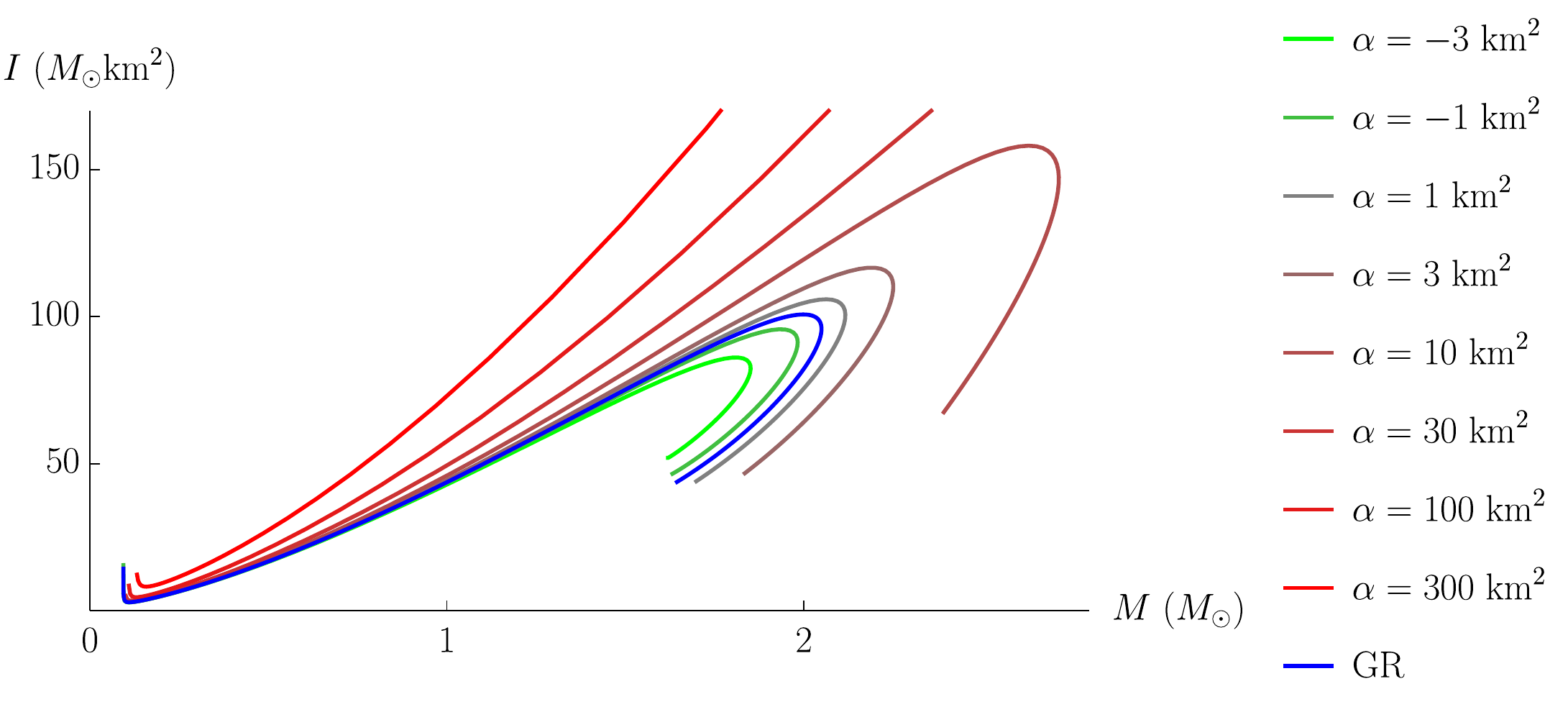}
\caption{Mass-inertia relations for the SLy EOS and various values of $\alpha$. The color code is the same as in Fig.~\ref{fig:MR}. For a given mass, a positive $\alpha$ leads to a larger moment of inertia, while a negative $\alpha$ has the opposite effect. This is consistent with Fig.~\ref{fig:MR}, which shows that the radius tends to increase with $\alpha$ for a given mass. 
}
\label{fig:IM}
\end{figure}

Let us now compute the moment of inertia $I$ of neutron stars. It is defined as the ratio between the angular momentum and the rotation speed, just like in classical mechanics: $I=J/\Omega$. To determine $I$, we integrate Eq.~\eqref{eq:tphi} numerically inside the star, starting from the center. The boundary conditions at the center are fixed by the expansion of this equation at the origin. Then, this solution is matched smoothly at the surface of the star with the exact exterior solution \eqref{eq:omegavac}. Some results are shown in Fig.~\ref{fig:IM} using the SLy EOS, for various choices of $\alpha$. 
Positive values of $\alpha$ (which are the only ones allowed for the SLy EOS, taking into account both the constraints from Secs.~\ref{sec:smallobj} and ~\ref{sec:statNS}) always increase the moment of inertia of the star for a given mass. Taking for instance $\alpha=10$~km$^2$, which is well in the range allowed by the results of Sec.~\ref{sec:statNS}, the moment of inertia already significantly deviates from the corresponding GR value (by as much as 20\% for the heaviest stars). For the anticipated measurement of the moment of inertia of the $1.338~M_\odot$ primary component of the double pulsar PSR J0737-3039A, the expected uncertainty (within GR) of $\sim 10\%$ is better than the deviation induced by a coupling constant of $\alpha \sim 40~{\rm km}^2$ (the rough upper bound on $\alpha$ we derive in Sec.~\ref{sec:statNS}).  Therefore, measurements of moments of inertia may become useful for constraining the theory of gravity, possibly strengthening the rough upper bound on $\alpha$ given in Sec.~\ref{sec:statNS}. They would in any case constrain $\alpha$ much better than weak field tests, by 2 or 3 orders of magnitude if we consider the best weak field test available, from the LAGEOS satellite \cite{Clifton:2020xhc}.

\section{Discussion}
\label{discussion}

In this paper, we have considered compact objects in a particularly interesting Horndeski model of gravity, which includes a representative of each one of the Horndeski terms. The theory is parametrized by a single coupling $\alpha$, which has the dimension of a length squared. This theory can be seen as originating from a higher-dimensional Einstein-Gauss-Bonnet theory (hence purely gravitational in higher dimensions). Remarkably, it possesses an exact vacuum solution, that can describe black holes. As one can naturally expect, deviations with respect to GR are suppressed by powers of $\alpha/l^2$, where $l$ is the characteristic curvature length. One can therefore follow two different strategies: either examine extremely precise tests using systems in the vicinity of the Earth, or situations where the precision is inferior but the curvature is stronger. We focused on the second strategy, considering black holes and neutron stars. 

A first compelling constraint in the black hole case practically excludes negative values of $\alpha$. Static star solutions allow us to establish independent constraints. Our analysis here only provides an estimate of these bounds, as some of the data we use was derived in the framework of GR. The exact bounds also depend on the details of the EOS of high-density matter and we examined several EOS that are compatible with current observational constraints. Our numerical solutions were constructed with two very different methods. Summarizing, assuming only a minimum black hole mass of $M=5.7^{+1.8}_{-2.1}~M_\odot$ for the lighter component of GW200115, yields a conservative estimate
\begin{equation}
0\leq\alpha\lesssim285^{+207}_{-171}~\text{km}^2
\end{equation}
(90\% credible interval).
This conservative estimate, can however be narrowed down under certain assumptions, typically to 
\begin{equation}
0\leq\alpha\lesssim59~\text{km}^2
\end{equation}
if the lighter component of GW190814 is a black hole, or even to 
\begin{equation}
0\leq\alpha\lesssim40~\text{km}^2
\end{equation}
if the GR estimates of neutron star radii in the BNS merger GW170817 are taken at face value (before correcting for the effects of deviations from GR).
This constitutes an improvement of at least one or two orders of magnitude with respect to the most stringent weak field tests \cite{Clifton:2020xhc}. A more detailed analysis should also take into account the rotation of neutron stars. We provided the first step in this direction, with the study of the slow rotation limit. Remarkably, we could integrate exactly the equations in the case of the vacuum solution. We also computed moments of inertia, which should soon be useful in light of binary pulsar data.

A key question that is still open is the perturbative stability of the solutions under consideration, both black holes and neutron stars. In particular, the theory could exhibit strong coupling issues for the scalar mode, as a standard kinetic term is absent from the action \eqref{eq:action}. Another interesting theoretical question is the analytic continuation of the solution \eqref{eq:vacf} beyond the horizon. By allowing the scalar field to depend on time, we provided a way to continue the solution inside the black hole, down to the center if $\alpha>0$. The metric, remarkably, is unaffected by this process. It displays an inner Cauchy horizon, and therefore an extremal limit; it is also more regular than the Schwarzschild solution at the origin. In a more observational perspective, detailed pulsar and binary black hole merger calculations are lacking, in particular a post-Newtonian treatment of the inspiral phase. The fact that the mass-radius curve of neutron stars seems to asymptote its black hole counterpart at high densities is a very remarkable feature of the theory we considered. A direct consequence is that the neutron star sequences all terminate at the same mass and radius, independently of the EOS. We are not aware of a similar behavior in other theories. We currently do not have an analytic explanation for this feature. We aim at investigating this aspect in detail in a future publication.

In the future, multi-messenger constraints coming from joint GW observations of BNS mergers and the associated electromagnetic counterparts, informed also by electromagnetic determination of neutron star radii, moment of inertia measurements of binary pulsars,  and laboratory experiments, may allow for the degeneracy between EOS effects and strong gravity effects on the neutron star mass-radius relation to be broken. This, in turn, will allow for new constraints on deviations from GR to be set. In this respect, the accurate determination of the threshold mass to black hole collapse of a BNS merger remnant \cite{2021PhRvD.103l3004B} will play a significant role, as well as whether a black hole formation is necessarily required in the short gamma ray burst mechanism (see discussion in \cite{2021GReGr..53...59S} and references therein).

\acknowledgments{We are grateful to Jolien Creighton for comments on the manuscript, and to Bhaskar Biswas, Tim Dietrich, Wolfgang Kaustaun, Panagiotis Iosif and George Pappas for useful discussions. We are also indebted to the anonymous referee for his insightful comments on the manuscript. This project has received funding from the European Union's Horizon 2020 research and innovation program under the Marie Sklodowska-Curie grant agreement No 101007855. 
A.L. thanks FCT for financial support through Project~No.~UIDB/00099/2020.
A.L. acknowledges financial support provided by FCT/Portugal through grants PTDC/MAT-APL/30043/2017 and PTDC/FIS-AST/7002/2020.
We also acknowledge  networking support by the GWverse COST Action
CA16104, ``Black holes, gravitational waves and fundamental physics.'' C.C. in particular thanks the Laboratory of Astronomy of AUTh in Thessaloniki for hospitality during the course of this work.
 N.S. acknowledges support by the COST actions CA16214 ``PHAROS" and CA18108 ``QG-MM". N.S. gratefully acknowledges the Italian Instituto Nazionale di Fisica Nucleare (INFN), the French Centre National de la Recherche Scientifique (CNRS) and the Netherlands Organization for Scientific Research, for the construction and operation of the Virgo detector and the creation and support of the EGO consortium.
}

\appendix

\section{First numerical scheme in Schwarzschild coordinates}
\label{sec:fieldeqs}

\begin{figure}[ht]
\centering
\includegraphics[scale=0.55]{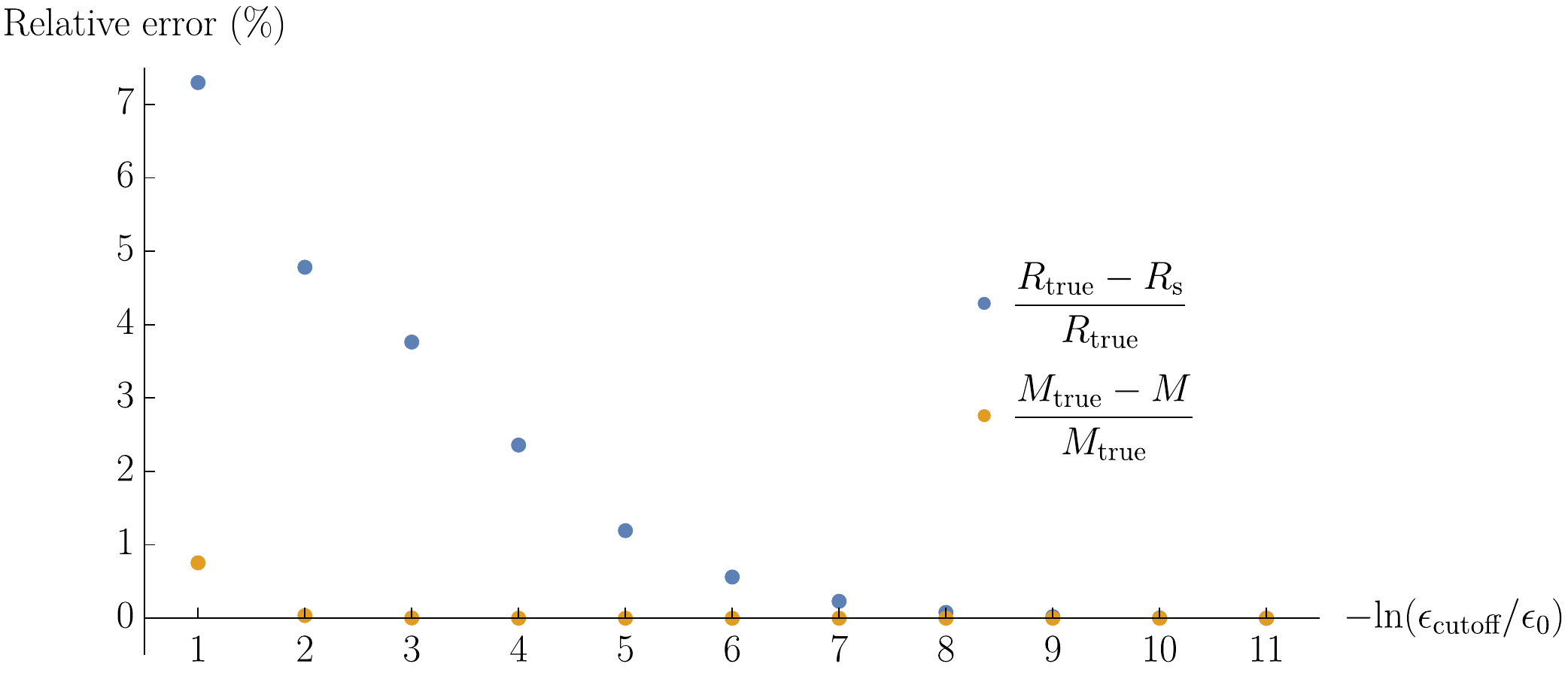}
\caption{Dependence of the radius and mass to the cutoff density, below which we stop the integration, for the method described in Sec.~\ref{sec:statNS}. Here, we solve the GR equations, using the SLy EOS, for a central density $\epsilon_0=10^{15}$~g/cm$^3$. The points show the relative difference between the radius (respectively mass) extracted using this cutoff density, and the most accurate value that we can compute, $R_\text{true}$ (respectively $M_\text{true}$). We see that the mass is pretty insensitive to the cutoff density, and stabilizes extremely fast. On the other hand, the error on the radius drops below $1\%$ only for $\epsilon_\text{cutoff}<\epsilon_0 10^{-6}$. In practice, we used a very low value of $\epsilon_\text{cutoff}=6.1\times10^5$~g/cm$^3$.}
\label{fig:epscutoff}
\end{figure}

To obtain a set of non-redundant field equations, we consider the $(tt)$ and $(rr)$ components of the metric field equations, together with the scalar field equation:
\begin{align}
\begin{split}
0&=f \left\{-4 \alpha  \left[(r \phi'-1)^2 f-1\right] \phi''+ \alpha \phi'^2 \left[\left(2-r^2\phi'^2\right) f+2\right]+1\right\}
\\
&\quad+f'\left\{2 \alpha \phi' \left\{1-\left[r \phi' (r \phi'-3)+3\right] f\right\}+ r\right\}-1+r^2 \kappa\epsilon,
\label{eq:tt}
\end{split}
\\
\begin{split}
0&=f h' \left\{2 \alpha  \phi' \left\{\left[r \phi' (r \phi'-3)+3\right] f-1\right\}- r\right\}
\\
&\quad+h \left\{f \left\{\alpha  \phi'^2 \left\{\left[r \phi' (8-3 r \phi')-6\right] f+2\right\}-1\right\}+1\right\}+r^2 h \kappa P,
\label{eq:rr}
\end{split}
\\
\begin{split}
0&=h \left\{2 f\left\{h' \left\{\phi'\left[(r \phi'-3) (r \phi'-1) f-1\right]-2 r (r \phi'-1) f \phi''\right\}\right.\right.
\\
&\quad\left.\left.+\left[1-(r \phi'-1)^2 f\right] h''\right\}+\left[1-3 (r \phi'-1)^2 f\right] f' h'\right\}+2\left\{\phi' \left[3 (r \phi'-1)^2 f-1\right] f'\right.
\\
&\quad\left.+2 f \left\{(r \phi'-1) f \left[(3 r \phi'-1) \phi''+2 \phi'^2\right]-\phi''\right\}\right\} h^2 +f \left[(r \phi'-1)^2 f-1\right] h'^2.
\label{eq:scaleq}
\end{split}
\end{align}
Thanks to shift symmetry, the scalar field equation \eqref{eq:scaleq} can be cast in the form of a current conservation equation, $\nabla_\mu J^\mu=0$. In spherical symmetry, the radial component of the current must then vanish, so Eq.~\eqref{eq:scaleq} boils down to 
\begin{equation}
0=\frac{4 \alpha  f}{r^2 h} \left[(r \phi'-1)^2 f-1\right] \left(2 \phi' h-h'\right).    
\label{eq:j_r}
\end{equation}
There are two branches of solutions to this equation. By continuity with the exterior solution, the branch we consider is the one for which $(r \phi'-1)^2 f=1$.
Additionally, the conservation of the matter stress-energy tensor, $\nabla_\mu T^{\mu\nu}=0$, translates as
\begin{equation}
0=\dfrac{h'}{2h} (P+\epsilon)+P'.
\label{eq:divT}
\end{equation}
The above system of first-order differential equations is solved as an initial value problem, up to the surface of the star, where it is matched to the analytic exterior solution. The numerical integration is terminated at a low cutoff density, just before the true surface. Figure \ref{fig:epscutoff} shows the influence of the cutoff density on the calculated mass and radius.

The small radius expansion \eqref{eq:fc}--\eqref{eq:Pc} is obtained by assuming that each function can be expanded in powers of $r$ close to the origin:
\begin{equation}
    F(r)\underset{r\to0}{=}\sum_{n=0}^\infty F_n r^n
\end{equation}
with $F=f$, $h$, $\phi$ $\epsilon$, $P$. This expansion is plugged in Eqs.~\eqref{eq:tt}--\eqref{eq:scaleq}-\eqref{eq:divT}, and solved order by order for the coefficients $F_n$. In order to match the interior solution with the exterior one, we require continuity of the metric and its first derivative at the surface. For the integration, we pick arbitrarily $h_0=1$ in the boundary condition $\eqref{eq:hc}$. We then obtain a numerical interior solution $h_\text{num}$ which needs to be rescaled to obtain the correct interior metric function, $h_\text{int}\equiv C h_\text{num}$, where $C$ is a constant. This constant is determined by requiring continuity of the metric function and its derivative:
\begin{align}
    h_\text{int}(R_\text{s})&=h_\text{ext}(R_\text{s}),
    \label{eq:hcont}
    \\
    h'_\text{int}(R_\text{s})&=h'_\text{ext}(R_\text{s}),
    \label{eq:dhcont}
\end{align}
where $h_\text{ext}$ is the solution \eqref{eq:vacf}. Equations \eqref{eq:hcont} and \eqref{eq:dhcont} fix the rescaling constant $C$, and the ADM mass of the solution $M$ that enters Eq.~\eqref{eq:vacf}.

\section{Second numerical scheme in isotropic coordinates}
\label{sec:altmethod}

The KEH/CST scheme \cite{10.1093/mnras/237.2.355,pub.1058503139} is a numerical method developed to obtain solutions for rotating relativistic stars and relies on the use of an isotropic radial coordinate $\scriptr$, on the integral representation of the metric potentials (making use of appropriate Green's functions) and on an iterative solution scheme, starting from an appropriate guess. Here, we briefly present the method's non-rotating limit, as applied to this specific Gauss-Bonnet gravity model. In practice, we use a dimensionless radial variable that compactifies infinity to a finite domain,\footnote{We use $s = \scriptr/(\scriptr+\scriptr_\text{e})$ as a new radial variable, where $\scriptr_\text{e}$ is the equatorial coordinate radius. The value of  $s = 0.5$ corresponds to the surface of the star, whereas $s = 1$ corresponds to spatial infinity. Note that, since we consider only spherically symmetric stars, $\scriptr_\text{e}$ is simply the (isotropic coordinate) radius at the surface of the star. The true circumferential radius is $R = e^{\mu_\mathrm{e}} \scriptr_\text{e} $, where $\mu_\mathrm{e}=\mu(\scriptr_\text{e})$, and $\mu$ is introduced in Eq.~\eqref{eq:isotrop}.} as introduced in \cite{pub.1058503139}. The full solution, both in the interior and in vacuum, out to infinity, is thus obtained simultaneously, without any matching at the surface. To our knowledge, this is the first application of the KEH/CST scheme to nonrotating stars in alternative theories of gravity.

The spacetime geometry of a spherically symmetric star in equilibrium using isotropic coordinates is described by the line element
\begin{equation}
\label{eq:isotrop}
\text{d}s^2 = -e^{2\nu}\text{d}t^2 + e^{2\mu}[\text{d}\scriptr^2 + \scriptr^2(\text{d}\theta^2 + \sin^2\theta \text{d}\varphi^2)],
\end{equation}
where $\nu(\scriptr)$ and $\mu(\scriptr)$ are metric functions. We derive the following components for the field equations of the theory (\ref{eq:action}), which are elliptic equations for $\nu$ and $\mu$: 
\begin{align}
\begin{split}
\nabla^{2}\nu = S_{\nu}(\scriptr) &\equiv e^{2\mu}\left\{4\pi e^{2\mu} (3p+\epsilon)\scriptr^2 - \scriptr \nu'\left[2+\scriptr(\mu'+\nu')\right] + \alpha e^{-2\mu}\left\{\phi'\left\{2\scriptr^2\mu^3\right.\right.\right.
\\ 
&\quad- 2\mu'^2(\nu'+\phi')\scriptr^2 - 2\scriptr\nu'^2(\scriptr\phi'-2) - \scriptr\left[\phi'^2(\scriptr\phi'-4)+4 \mu''\right]
\\
&\quad+ 2\mu'\left\{-2 + \scriptr\left[2\scriptr\nu'^2 +\nu'(4-\scriptr\phi')+\phi'(-2+\scriptr\phi')-2\scriptr\mu''\right]\right\}
\\
&\quad+ 2\nu'\left.\left\{2+\scriptr\left[\phi'(\scriptr\phi'-4) + 2\scriptr\mu''\right]\right\}\right\} + 2\scriptr\left\{-\scriptr\mu'^2\right.+2\mu'(-1+\scriptr\nu')
\\
&\quad+\left.\left.\left. \scriptr\left[\phi'^2 + \nu'(2-2\scriptr\phi')\right]\phi''\right\}\right\}\right\}/\left\{\scriptr\left[e^{2\mu} \scriptr + 2\alpha\phi'(-2-2\scriptr\mu'+\scriptr\phi')\right]\right\} 
\\
&\quad+ \dfrac{2\nu'}{\scriptr},
\label{eq:source_nu}
\end{split}
\\
\begin{split}
\nabla^{2}\mu = S_{\mu}(\scriptr) &\equiv -\left\{8\pi e^{2\mu}\epsilon \scriptr^2 +e^{2\mu}\scriptr\mu'(4+\scriptr\mu')+ \alpha(2\mu' - \phi')\phi'\left[-4 + \scriptr^2(2\mu'^2\right.\right.
\\
&\quad- 2\mu'\phi'\left. + \phi'^2)\right]- 4\alpha \scriptr\left.(\mu' - \phi')(2 + \scriptr\mu' - \scriptr\phi')\phi''\right\}
\\
&\quad/\left\{2\scriptr\left[e^{2\mu}\scriptr + 2\alpha\phi'(-2-2\scriptr\mu'+\scriptr\phi')\right]\right\}
\\ 
&\quad+ \dfrac{2\mu'}{\scriptr},
\label{eq:source_mu}
\end{split}
\end{align}
where $\nabla^2 = \partial_{\scriptr\scriptr} + \frac{2}{\scriptr}\partial_\scriptr$  is the flat-space Laplacian and $S_{\nu}$, $S_{\mu}$  are the source terms of the elliptic equations.  Using the Green's function of the three-dimensional Laplacian operator in spherical coordinates, the elliptic field equations can be cast into the following integral equations \cite{10.1093/mnras/237.2.355}:

\begin{align}
   \nu  &= -\frac{1}{4\pi}\int_{0}^{\infty}\text{d}\scriptr'\int_{0}^{\pi}\text{d}\theta'\sin\theta'\int_{0}^{2\pi}\text{d}\varphi'\scriptr^{'2}\frac{S_{\nu}(\scriptr')}{|\scriptr-\scriptr'|}.
    \label{eq:nu}
\\
 \mu &= -\frac{1}{4\pi}\int_{0}^{\infty}\text{d}\scriptr'\int_{0}^{\pi}\text{d}\theta'\sin\theta'\int_{0}^{2\pi}\text{d}\varphi'\scriptr^{'2}\frac{S_{\mu}(\scriptr')}{|\scriptr-\scriptr'|} \label{eq:mu},
\end{align}
In isotropic coordinates, the branch of solutions to the current equation \eqref{eq:j_r} that we have chosen (and that yields physically reasonable solutions) is written as:
\begin{equation}
    \phi' = \mu'.
    \label{eq:phi}
\end{equation}
The hydrostationary equilibrium equation \cite{friedman_stergioulas_2013} in the non-rotating limit can be written as:

\begin{equation}
    \nabla_\alpha(H-\text{ln} \:u^t) = 0
    \label{eq:hydro}
\end{equation}
where $u^t = e^{-\nu}$ and $H$ is the specific enthalpy, which is defined as
\begin{equation}
    H(P) = \int_{0}^{P}\frac{\text{d}P^\prime}{\epsilon(P^\prime) + P^\prime}\,.
    \label{eq:spec_enthalpy}
\end{equation}
The iterative procedure can be summarized in the following steps, during which the central energy density ${\epsilon}_0$ is held fixed (see \citep{friedman_stergioulas_2013} for more details):

\begin{enumerate}
    \item The iteration starts with an initial guess\footnote{We use a Tolman-Oppenheimer-Volkoff solution in GR, with the same central density, as an initial guess. The scalar field is initialized through  \eqref{eq:phi}.} for the metric potentials  $\nu$, $\mu$, the scalar field $\phi$, the energy density $\epsilon$ and the  coordinate radius of the surface $\scriptr_\text{e}$.
    \item The metric potentials and scalar field are rescaled by $\scriptr_\text{e}^2$. 
    \item Equating the first integral of the hydrostationary equilibrium~\eqref{eq:hydro} at the center to its value at the surface,  we obtain a new value of the equatorial coordinate radius $\scriptr_\text{e}$.
    \item Using the newly obtained $\scriptr_\text{e}$, we calculate new distributions for the enthalpy, pressure and energy density, equating the first integral of the hydrostationary equilibrium at each point inside the star to its value  at the surface.
    \item New distributions of the metric potentials and scalar field are obtained from~\eqref{eq:mu},~\eqref{eq:nu},\\~\eqref{eq:phi} using the new value of $\scriptr_\text{e}$ and the enthalpy distribution calculated in the previous step. 
\end{enumerate}
The above procedure is repeated starting from the second step until convergence is achieved with the desired accuracy. In step 2, the metric potentials and scalar field are rescaled in the following way:
\begin{align}
\hat\mu = \frac{\mu}{\scriptr_\text{e}^2},\qquad \hat\nu= \frac{\nu}{\scriptr_\text{e}^2}, \qquad \hat\phi = \frac{\phi}{\scriptr_\text{e}^2}.
\end{align}
Calculating the first integral of the hydrostationary equilibrium at the location of the maximum energy density and equating with its value at the surface:
\begin{equation}
    H_\text{max} - \text{ln}\: u^t_\text{max} = H_\text{surface} - \text{ln}\: u^{t}_\text{surface},
\end{equation}
we obtain the following equation for the equatorial radius (step 3)
\begin{equation}
    \scriptr_\text{e}^2 = \frac{H_\text{max}}{\hat{\nu}_\text{surface}-\hat{\nu}_\text{max}}\,.
\end{equation}
In order to obtain a new distribution for the enthalpy, we solve the first integral of the hydrostationary equilibrium and equate its value at an arbitrary point inside the star to its value at the surface (step 4):
\begin{equation}
    H = H_\text{surface} + \text{ln}\:\frac{u^t}{u^t_\text{surface}}\,.
\end{equation}
In our implementation of the KEH/CST scheme the first-order derivatives are approximated by a standard three-point formula. Using the same formula for the second-order derivatives, however, introduces a significant oscillation of the metric potential $\mu$. This problem was resolved\footnote{An alternative solution to this problem was first introduced in \cite{Stergioulas_1995}.} by approximating the second-order derivatives with two consecutive first-order derivatives evaluated by a standard three-point formula and applying a Savitzky–Golay filter \cite{doi:10.1021/ac60214a047} which removed the oscillations. As a variable for monitoring the convergence of this method we chose the surface radius of the star $\scriptr_e$. We terminate the iteration, when the relative difference in $\scriptr_e$ between two consecutive iterations drops below $10^{-8}$. Additionally, in order to improve\footnote{This was found to be critical for the convergence of the method for large values of the coupling constant $\alpha$.} the convergence of the method, we employed the under-relaxation technique \cite{10.1093/mnras/237.2.355}.

\begin{figure}[ht]
\centering
\includegraphics[scale=0.45]{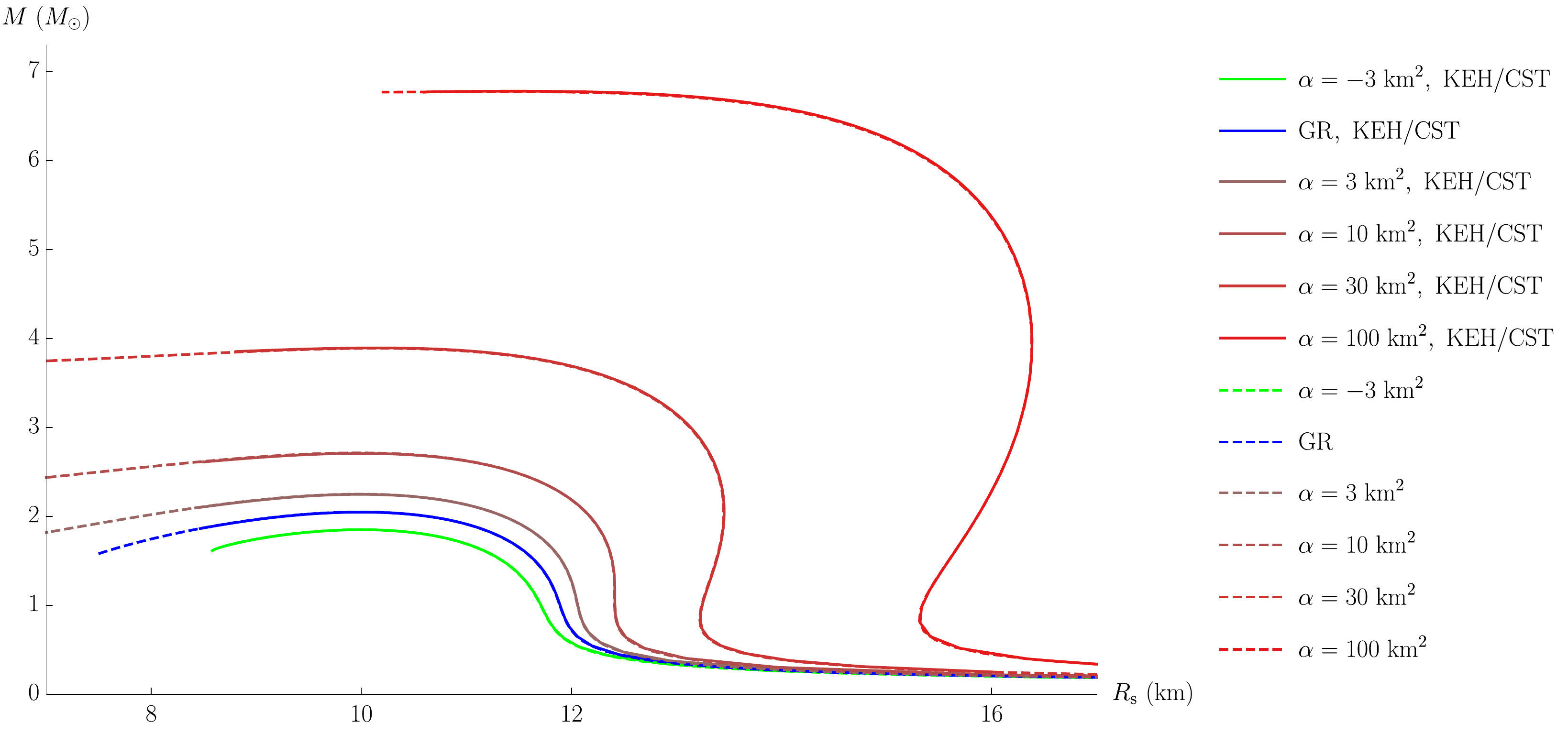}
\caption{Comparison of the mass-radius relations for the SLy EOS and different values of $\alpha$, obtained with the two different numerical methods. The results are practically identical, with a maximum relative deviation in the order of $0.01\%$.}
\label{fig:MR_comparing}
\end{figure}

In Fig.~\ref{fig:MR_comparing}, we compare mass-radius relations for the SLy EOS and different values of $\alpha$, obtained with the two different numerical methods. The results are almost identical, with a maximum relative deviation in the order of $0.01\%$.

\bibliographystyle{unsrt}
\bibliography{biblio}
\end{document}